\documentclass[pre,twocolumn,superscriptaddress,showpacs]{revtex4}
\usepackage{graphicx}
\usepackage{amssymb}
\usepackage{amsmath}

\begin{document}

\title{Bayesian decision making in human collectives with binary choices}
\author{V\'ictor M. Egu\'iluz}
\email{victor@ifisc.uib-csic.es}
\affiliation{Instituto de F\'isica Interdisciplinar y Sistemas Complejos IFISC (CSIC-UIB), E07122 Palma de Mallorca, Spain}
\author{Naoki Masuda}
\affiliation{Department of Engineering Mathematics, University of Bristol, Merchant Venturers Building, Woodland Road, Clifton, Bristol BS8 1UB, United Kingdom}
\author{Juan Fern\'andez-Gracia}
\affiliation{Instituto de F\'isica Interdisciplinar y Sistemas Complejos IFISC (CSIC-UIB), E07122 Palma de Mallorca, Spain}
\affiliation{Instituto Mediterr\'aneo de Estudios Avanzados IMEDEA (CSIC-UIB), E07190 Esporles, Spain}

\date{\today}

\begin{abstract}
Here we focus on the description of the mechanisms behind the process of information aggregation and decision making, a basic step to understand emergent phenomena in society, such as trends, information spreading or the wisdom of crowds. In many situations, agents choose between discrete options. We analyze experimental data on binary opinion choices in humans. The data consists of two separate experiments in which humans answer questions with a binary response, where one is correct and the other is incorrect. The questions are answered without and with information on the answers of some previous participants. We find that a Bayesian approach captures the probability of choosing one of the answers. The influence of peers is uncorrelated with the difficulty of the question. The data is inconsistent with Weber's law, which states that the probability of choosing an option depends on the proportion of previous answers choosing that option and not on the total number of those answers. Last, the present Bayesian model fits reasonably well to the data as compared to some other previously proposed functions although the latter sometime perform slightly better than the Bayesian model. The asset of the present model is the simplicity and mechanistic explanation of the behavior.
\end{abstract}

\maketitle

%%%%%%%%
\section{Introduction}
The process of information aggregation in social systems gives rise to emergent phenomena like the wisdom of crowds \cite{Galton1907,Surowiecki2005}. In order to understand such phenomena a quantitative understanding of the mechanisms by which information is aggregated and used in opinion formation and decision making is needed. In the case of the wisdom of crowds, which refers to having a better estimation of the solution to a question when the opinions of multiple heterogeneous agents are aggregated, it has been shown that social interaction can lead to misleading estimations \cite{Lorenz2011}. The issue of information aggregation is a hot topic which is expected to give insights into the solution of many societal problems. For example, 2014's World Economic Forum's meeting has the title ``Leveraging collective intelligence for
unprecedented challenges''.

Models of opinion dynamics are based on assumptions on the decision
making process on interacting individuals.
Simple decision making rules employed in these
models include proportional imitation (i.e., the rate of the opinion
conversion is proportional to the number of peers possessing the
different opinion), employed in the voter model, majority rules (i.e.,
the same rate is a superlinear function),
thresholding rules (i.e., thresholding function), reinforcement rules (i.e., adaptive function depending on experiences of agents), and homophily rules (i.e., similar individuals more likely interact) \cite{Axelrod1997book,Castellano2009,Galam2012book}. The type of the employed decision making rule affects the
possibility, final state, speed, and other dynamical phenomena of collective opinion
formation. However, in
physics and even social sciences literature,
justification of these different types of models
is at best based on a qualitative assessment of human behavior. Beyond opinion dynamics, social dilemmas, which in many cases are based on binary decision making, also offer an opportunity to bridge theory to experiments \cite{Helbing2010,Szolnoki2012}.

For animals in groups,
recent work in this direction has identified Bayesian inference as a mechanism behind their collective behavior \cite{McNamara1980,Kahneman1982,McNamara2006Oikos,Martins2009PhysicaA,McKay2010,Trimmer2011,Perna2012PlosComputBiol,Andreoni2012AmEconJMicroecon,Arganda2012PNAS,Johnson2013,Marshall2013,Marshall2013b,Perez-Escudero2013}.
In humans, experimental evidence of Bayesian inference has been provided in the realm of
perceptual and cognitive domains \cite{KnillPouget2004TrendsNeurosci,Tenenbaum2011Science}.
Effects of Bayesian types of inference on collective behavior have been investigated with the use of mathematical
and individual-based models
\cite{Banerjee1992QuarterlyJEcon,Orlean1995JEconBehavOrgan,Martins2008IntJModPhysC,Binmore2008book,AcemogluOzdaglar2011DynGamesAppl,Andreoni2012AmEconJMicroecon,Nishi2013PhysRevE}.
Toward quantitative understanding of social decision making of humans,
the seminal experiment by Milgram and colleagues \cite{Milgram1969} designed to assess the probability to stop by a group of bystanders has recently been reproduced \cite{Gallup2012} whose results are fitted by a heuristic function.
There are also other recent studies
attempting to fit Bayesian (see the references above), evolutionary dynamical
\cite{Traulsen2010PNAS_exper},
and other \cite{Mori2012PhysRevE,Mori2013JPhysSocJapan} models
to behavioral data.
The wisdom of crowds when interaction among participants is allowed is also a target of recent experimental studies \cite{Lorenz2011,King2012,Moussaid2013}.
However, a unifying quantitative framework to infer models of social decision making
on the basis of
behavioral data of humans is still lacking and much preceded by
accumulating modeling frameworks for social animals
\cite{McNamara2006Oikos,Perna2012PlosComputBiol,Arganda2012PNAS}.

%%%%%%%%%%%%%%%%%%%%%%%%%%%%%%%%Fig 1
\begin{figure}[ht]
\centerline{\includegraphics[width=0.45\textwidth]{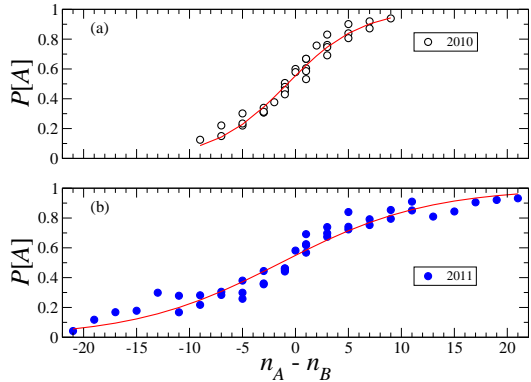}}
\caption{{\bf Bayesian inference and experimental data.} We plot the probability to report a correct answer $A$ as a function of $n_A-n_B$ for various $(n_A, n_B)$ pairs: (a) Data set $D_1$, (b) Data set $D_2$. The circles correspond to the data. The solid curves indicate the best fits of Eq.~\eqref{eq:final}: $(p,s)=(0.80,0.75)$ in (a) and $(0.82,0.87)$ in (b).}
\label{f:aggregate}
\end{figure}

In the present study, we address the potential of the Bayesian approach
to explain human decision making under social interaction.
We focus on subjects answering questions with binary options, one of which is correct. This situation contrasts with that of the previous studies on the wisdom of crowds that allowed virtually real values of answers \cite{Lorenz2011,King2012,Moussaid2013}. We examine binary choices because
many options in nature are discrete, as exemplified by voting, purchasing, and
deciding where to live. In many of such situations, extrapolation from continuous settings is not obvious.
We use previously published data sets in which the participants first answer in the absence of social information and later with the information about the answers submitted by the $r$ previous respondents; $r$ gradually increases for the same question \cite{Mori2012PhysRevE,Mori2013JPhysSocJapan}. The participants answer in a sequence, the situation akin to that
for previous Bayesian models of the emergence of herd behavior
\cite{Bikhchandani1992JPolitEcon,Banerjee1992QuarterlyJEcon}.
We show that simple Bayesian models reasonably explain the behavioral data.

\section{Materials and Methods}

\subsection{Model}

We denote the two options of a question by $A$ and $B$
Without loss of generality, we assume that $A$ and $B$ are the correct and wrong answers of the question $q$, respectively.
We label the $N$ agents $1,\ldots,N$ and denote the option that agent $i$ ($i=1,\ldots,N$) selects in question $q$ by
$x_i(q) \in \left\{ A, B \right\}$. We denote by $P\left[x_i(q)=A\right]$ the strength of the belief (hereafter, simply the belief), with which agent $i$ believes in $A$. A parallel definition is applied to $P\left[x_i(q)=B\right]$. Note that $P\left[x_i(q)=A\right], P\left[x_i(q)=B\right] \ge 0$, and $P\left[x_i(q)=A\right] + P\left[x_i(q)=B\right] = 1$.

We update the agent $i$'s belief as follows.
We assume that the answer of the previous respondent $j$, i.e., $x_j(q)$, is generated according to the probability specified by the belief of agent $j$, i.e., $P\left[x_j(q)=A\right]$, which equals $1 - P\left[x_j(q)=B\right]$.
Then, by using the Bayes' theorem, agent $i$ is assumed to update the
belief on the basis of the old belief and $x_j(q)$.
The posterior belief of agent $i$ is given by
\begin{align}
P&\left[x_i(q)=A\right]_{\rm post}\notag\\
=&\frac{P\left[x_j(q)=A | x_i(q)=A\right] P\left[x_i(q)=A\right]_{\rm pre}}
{\sum_{X_i=A, B}P\left[x_j(q)=A | x_i(q)=X_i\right] P\left[x_i(q)=X_i\right]_{\rm pre}}\notag\\
=&
\frac{c P\left[x_i(q)=A\right]_{\rm pre}}{c P\left[x_i(q)=A\right]_{\rm pre} + (1-c) P\left[x_i(q)=B\right]_{\rm pre}},
\label{eq:PiA_update}
\end{align}
where $P\left[x_i(q)=A\right]_{\rm pre}$ and
$P\left[x_i(q)=B\right]_{\rm pre}$ are prior beliefs summing up to unity.
Parameter $c \equiv P\left[x_j(q)=A | x_i(q)=A\right]$ ($1/2\le c < 1$)
represents the flexibility of agent $i$ in response to agent $j$'s answer.
If $c$ is close to unity, $P\left[x_j(q)=B | x_i(q)=A\right] = 1 - c$
is small such that
$1-P\left[x_i(q)=A\right]_{\rm post}$, i.e.,
$P\left[x_i(q)=B\right]_{\rm post}$ is large once agent $i$ observes $x_j(q)=B$ for a given
$P\left[x_i(q)=A\right]_{\rm pre}$. If $c$ is close to $1/2$,
$P\left[x_i(q)=A\right]_{\rm post}$ is insensitive to $x_j(q)$. By symmetry, we assumed that
$P\left[x_j(q)=B|x_i(q)=B\right]=c$ such that $P\left[x_j(q)=A|x_i(q)=B\right]= 1-P\left[x_j(q)=B|x_i(q)=B\right]=1-c$.

Iterative application of Eq.~\eqref{eq:PiA_update} leads to
\begin{align}
P&\left[x_i(q)=A\right]\notag\\
=&\frac{c^{n_A-n_B} P_0\left[x_i(q)=A\right]}{c^{n_A-n_B} P_0\left[x_i(q)=A\right]
+ (1-c)^{n_A-n_B} \left\{1-P_0\left[x_i(q)=A\right]\right\}}
\label{eq:PiA_specific_form}
\end{align}
and $P\left[x_i(q)=B\right] = 1-P\left[x_i(q)=A\right]$,
where $n_A$ and $n_B$ are the
accumulated
numbers of $A$ and $B$ responses of the previous respondents
observed by agent $i$, respectively.
The initial belief of agent $i$ in option $A$ is denoted by
$P_0\left[x_i(q)=A\right]$.
It should be noted that the order in which the previous responses are observed does not affect $i$'s behavior.
The belief of each agent $i$ is uniquely determined by $n_A-n_B$ and the initial belief. We can rewrite Eq.~\eqref{eq:PiA_specific_form} as
\begin{equation}
P\left[x_i(q)=A\right]=\frac{1}{1+ps^{n_A-n_B}}~,
~\label{eq:final}
\end{equation}
where $p=\left\{1-P_0\left[x_i(q)=A\right]\right\}/P_0\left[x_i(q)=A\right]$ and
$s=(1-c)/c$. Previous studies used Eq.~\eqref{eq:final} to account for consensus decision making by fish \cite{Ward2011PNAS,Perez-Escudero2011PlosComputBiol}.

%%%%%%%%%%%%%%%%%%%%%%%%%%
\subsection{Data set}\label{sec:data}

In the present study, we use the
two data sets collected in Refs.~\cite{Mori2012PhysRevE,Mori2013JPhysSocJapan}.
The first data set, which we denote by $D_1$,
consists of two sets of face-to-face
experiments \cite{Mori2012PhysRevE}.
Data set $D_1$ consists of the results obtained from two populations of subjects
each of which contains $N=31$ subjects (KUE-A and KUE-B in Ref.~\cite{Mori2012PhysRevE}).
Each subject
went through 100 questions. Each question allowed binary options, one being correct and the other being incorrect.
Generally speaking, the subjects were asked to answer each question
more than once under different information conditions. We refer to a sequence of answering sessions under a given question $q$ ($1\le q\le 100$)
and information condition parameterized by $r$ as a round. Subjects went through several rounds for each question in general.

%%%%%%%%%%%%%%%%%%%%%%%%%%%%%%%%%%Fig 2
\begin{figure}[t]
\centerline{\includegraphics[width=0.45\textwidth]{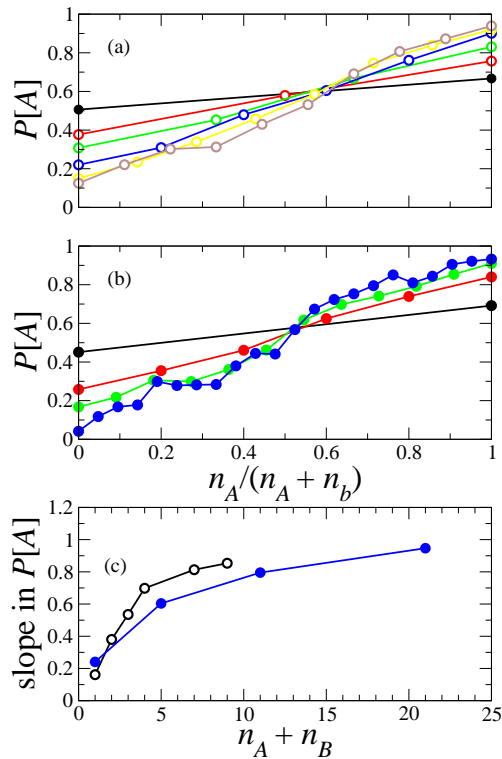}}
\caption{{\bf Dependence on the fraction of correct answers.} (a) Probability to answer correctly as a function of the
  fraction of correct answers of the previous respondents for data set $D_1$.
Black $r=1$, red $r=2$, green $r=3$, blue $r=4$, yellow $r=7$, and brown $r=9$.
(b) Same results for data set $D_2$.
Black $r=1$, red $r=5$, green $r=11$, and blue $r=21$.
(c) Slope as a function of $r$
obtained by the least square method applied to the plots
in panels (a) and (b). The closed and open circles correspond to
$D_1$ and $D_2$, respectively.}
\label{f:voterlike}
\end{figure}

The number of rounds that a subject experienced for each question depends on the subject. The $N$ subjects in a population
were randomly assigned labels 1, 2, $\ldots$, $N$.
In the first round, all subjects answered the question without referring to others' responses. This is the memoryless condition ($r=0$).
If everybody answered within the allocated time,
there were $N$ data points for each population and question.

The second round was implemented as follows.
First, subject 1 left this question
without participating in the second and following rounds.
Second, subject 2 observed the answer of subject 1 in the first round and possibly updated the private answer. Similarly, subject $i$ observed subject $(i-1)$'s answer in the first round and possibly updated the answer, where
$i$ runs from $i=3$ to $i=N$ in an ascending order.
In the best case whereby everybody answered,
$N-1$ data points were collected in the second round.
The collected data correspond to information condition $r=1$.

The third round, corresponding to $r=2$, was implemented as follows.
First, subject 2 left without participating in the third and
further rounds. Second, subject 3 observed the number of answers
($n_A$, $n_B$) submitted most recently by the previous $r=2$
respondents and answered the question again. It should be noted that
($n_A$, $n_B$)= (2, 0), (1, 1), or (0, 2). To calculate ($n_A$,
$n_B$), the answer of subject 1 in the first round and that of subject
2 in the second round were used. This is because subject 1 already
left the question before the second round. In other words, the answer
of subject 1 is assumed to be quenched to that made in the first round
in the subsequent (i.e., second and later) rounds. Third, subject $i$
answered after observing ($n_A$, $n_B$) calculated on the basis of the
most recent choice of subjects $i-1$ and $i-2$, where $i$ runs from 4
to $N$. There are at most $N-2$ answers obtained from the third round.

After the third round was completed,
further rounds were carried out with
$r=3$, 5, 7, 9, and $\infty$ in this order, where $r=\infty$ implies that the subjects can refer to the most recent answers of all the preceding respondents.
Subject 3 had left before the fourth round, corresponding to $r=3$, started.
Subjects 4 and 5 had left before the fifth round, corresponding to $r=5$, started. There are eight rounds in total.
The labels of the subjects were fixed throughout the 100 questions.

The second data set, which we denote by $D_2$, consists of two
sets of web-based experiments. They are denoted by
HUE-A and HUE-B in Ref.~\cite{Mori2012PhysRevE} and
the O and C treatments, corresponding to
$r=0$ and $r>0$, respectively, in Exp-II in Ref.~\cite{Mori2013JPhysSocJapan}.
Data set $D_2$ consists of the results obtained from two subject populations each of which contains $N=52$ subjects. Each subpopulation of subjects went through 120 questions. In $D_2$, each subject experienced up to 6 rounds, i.e.,
$r=0, 1, 5, 11, 21$, and $\infty$ for each question. The labels of the subjects were randomly shuffled in the beginning of each question.

% Results and Discussion can be combined.
\section{Results}

\begin{figure}[t]
\centering
\includegraphics[width=0.4\textwidth]{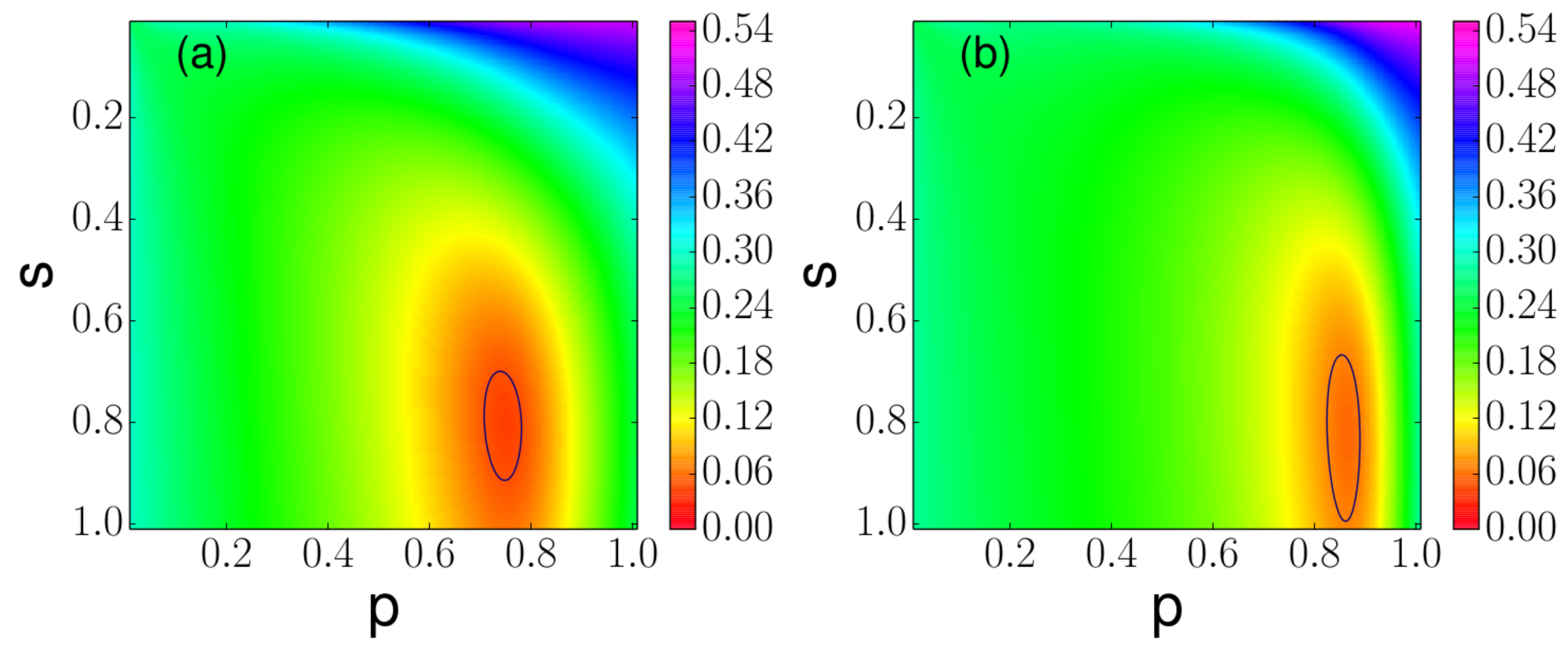}
\caption{{\bf Parameter estimation.} Root mean squared error associated to the fitting of the model given by Eq.~{eq:final} to datasets (a) $D_1$ and (b) $D_2$. The contour line shows a level of $0.05$ and $0.07$ in (a) and (b), respectively.}
\label{f:fittingerr}
\end{figure}

Let us first consider
the aggregate results for each experiment.
As described previously, a subject answers a question
after observing the number of the correct answer, $n_A$,
and that of the incorrect answer, $n_B$,
from the last $r=n_A + n_B$ respondents.
By the aggregate results we mean that we aggregate the number of correct answers across questions for the same condition $(n_A, n_B)$.
We denote by $R(n_A, n_B)$ the number of
answers obtained under condition ($n_A$, $n_B$), summed over respondents
$i$ and questions $q$.
Out of these answers,
the number of answer $A$, denoted by $N_A(n_A, n_B)$, is given by
\begin{equation}
N_A(n_A, n_B) = \sum_{i, q} x_i(q, n_A, n_B).
\end{equation}
The fraction of $A$ answers under condition ($n_A$, $n_B$) is given by
$N_A(n_A, n_B)/R(n_A, n_B)$. This fraction for various ($n_A$,
$n_B$) pairs is plotted in Fig.~\ref{f:aggregate}(a) and ~\ref{f:aggregate}(b) for $D_1$ and $D_2$, respectively. We fit
$P[x_i=A]$ given by Eq.~\eqref{eq:final} to the experimental data, where we suppress $q$ in the argument of $x_i$ because we have aggregated the data over the questions.
We estimate the values of $p$ and $s$ by an exhaustive sampling in the parameter space. For each sampled $(p,s)$ pair, we calculate the error by the total square distance between Eq.~\eqref{eq:final} and the empirical values summed
over the available $(n_A, n_B)$ pairs.
The parameter values yielding the smallest error are adopted.
The results of the best fitting are
shown by the solid curves
in Fig.~\ref{f:aggregate}(a) and \ref{f:aggregate}(b) for
$D_1$ and $D_2$, respectively.
For $D_1$, the best fit is obtained for
$p=0.81$, $s=0.75$ which lead to a root mean squared error RMSE$\approx 0.042$.
For $D_2$, we obtain $p=0.82$, $s=0.87$ leading to RMSE$\approx0.059$.
Figure~\ref{f:aggregate} indicates that
Eq.~\eqref{eq:final} fits both data sets
reasonably well. The value of the RMSE as a function of both parameters is shown in Fig.~\ref{f:fittingerr}.

%%%%%%%%%%%%%%%%%%%%%%%%%%%%%%%%Fig 3
\begin{figure}[t]
\centerline{\includegraphics[width=0.45\textwidth]{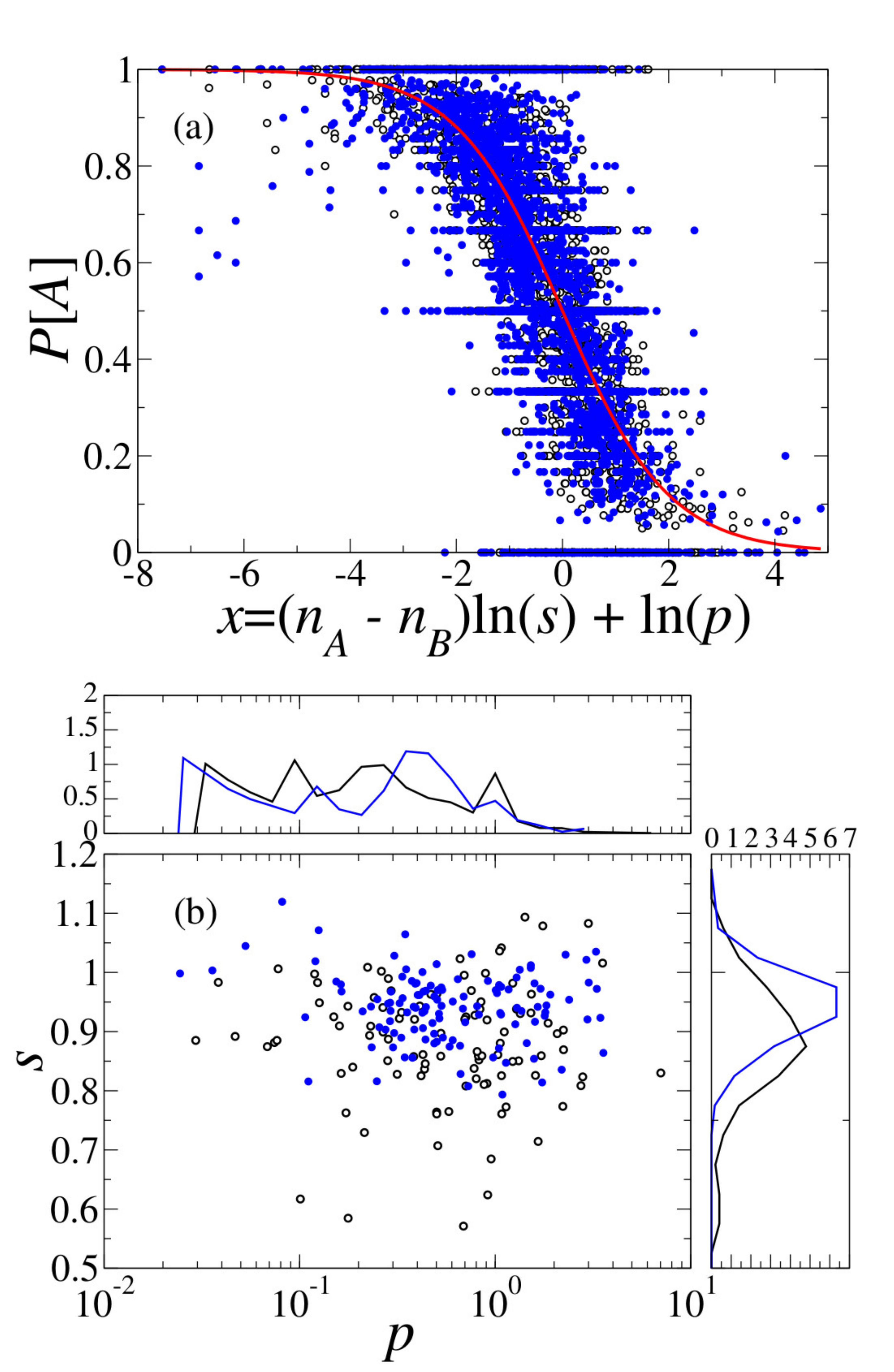}}
%\centerline{\includegraphics[width=0.45\textwidth]{Fig3b}}
\caption{{\bf Dependence on the question.} (a) Probability of correct answers as function of rescaled accumulated answers of previous respondents $(n_A-n_B)\ln(s) + \ln(p)$. Each symbol represents a question. Black, open symbols correspond to $D_1$, and blue filled circles to $D_2$. We estimated the $s$ and $p$ values for each question by applying the least square method to the data for the corresponding question. (b) Estimated $p$ and $s$ values for different questions. The top and side panels show the distributions of $p$ and $s$, respectively.
\label{f:allps}}
\end{figure}

An alternative hypothesis of collective decision making
is that $P[x_i=A]$ obeys Weber's law such that it is a function that only depends on
$(n_A-n_B)/(n_A+n_B)$, or equivalently, $n_A/(n_A+n_B)$ \cite{Perna2012PlosComputBiol,Arganda2012PNAS}.
To test this hypothesis, we aggregate the data over $q$ and $i$
using the same aggregation as that used in Fig.~\ref{f:aggregate}, but separately for $r$ to examine the effect of $r$ on the decision making,
and plot $P[x_i=A]$ as a function of $n_A/(n_A+n_B)$. The results
are shown in Fig.~\ref{f:voterlike}(a) and \ref{f:voterlike}(b) for $D_1$ and $D_2$, respectively. Each color corresponds to a value of $r=n_A + n_B$. If Weber's law holds true, all curves collapse on a single curve. Figure~\ref{f:voterlike} indicates that it is not the case.
To be more quantitative,
in Fig.~\ref{f:voterlike}(c),
we plot the slope of the curves obtained by applying the least square method to the data shown in Fig.~\ref{f:voterlike}(a) and \ref{f:voterlike}(b).
The figure indicates that the slope increases with $r (=n_A+n_B)$ and seems to saturate. That would mean that Weber's law is correct for sufficiently large $r$ values. Nevertheless, for the $r$ values accessed by the experiment, Weber's law does not hold. With data for larger $r$ values one could assess if Weber's law holds and from which $r$ value on.

We have a reasonable fit of the data to Eq.~\eqref{eq:final} even without aggregation over the questions. To show this, for a given question, we calculate
the fraction of the correct answers
$N_A^q(n_A, n_B)/R^q(n_A, n_B)$, where
$R^q(n_A, n_B)$ is the number of answers to question $q$ obtained under condition $(n_A, n_B)$, and
$N_A^q (n_A, n_B) \equiv \sum_i x_i(q,n_A, n_B)$ is the corresponding
number of answer $A$. The relationship between $P\left[x_i(q)=A\right]$ and
$z = (n_A-n_B)\ln s_q + \ln p_q$ for different questions is plotted in
Fig.~\ref{f:allps}(a). If Eq.~\eqref{eq:final} holds true,
the results for different questions should collapse on a single curve
$P\left[x_i(q)=A\right]=\left[ 1+\exp(z)\right]^{-1}$ shown by the solid line. The results for the different questions do roughly collapse on this curve.
The estimated values of $p_q$ and $s_q$ for individual questions are shown in
Fig.~\ref{f:allps}(b). As before, we obtained parameter values $p_q$ and $s_q$ by sampling the parameter space and finding the values giving the smallest error. Figure~\ref{f:allps}(b) shows that the estimated parameter values depend on the question to a large extent. For some questions, $p>1$, implying that
the initial belief in the correct answer is worse than the
random coin flip, i.e., $P_0\left[x_i(q)=A\right] < 0.5$. For a majority of questions, however, the initial belief is better than the random coin flip, and for some questions, it is quite accurate (for example, $p=0.1$ corresponds to $P_0\left[x_i(q)=A\right]=0.91$). Another remark is that $p$ and $s$ are apparently uncorrelated. This implies that the flexibility of the opinion change does not depend on the difficulty of the question.

%%%%%%%%%%%%%%%%%%%%%%%%%%%%%%%%Fig 4
 \begin{figure}[!ht]
\centerline{\includegraphics[width=0.5\textwidth]{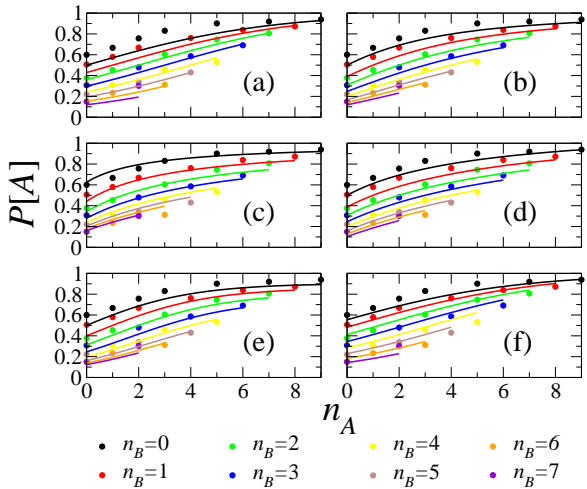}}
  \caption{
    {\bf Testing different models for data set $D_1$.}
    (a) $P[A]=\delta^{n_A}/(\delta^{n_A}+\delta^{n_B})$, $\delta=1.33$, RMSE $=0.06$,
    (b) $P[A]=(\delta+n_A)^{\epsilon}/[(\delta+n_A)^{\epsilon}+(\delta+n_B)^{\epsilon}]$, $\delta=3.90$, $\epsilon=1.95$, RMSE $=0.05$,
    (c) $P[A]=(\delta+\epsilon n_A)/[1+\epsilon(n_A+n_B)]$, $\delta=0.63$, $\epsilon=0.41$, RMSE $=0.03$,
    (d) $P[A]=1/2+\delta(n_A-n_B)/(n_A+n_B+\epsilon)$, $\delta=0.68$, $\epsilon=5.06$, RMSE $=0.06$,
    (e) $P[A]=[1+(1+\delta\epsilon^{-n_A})/(1+\delta\epsilon^{-n_B})]^{-1}$, $\delta=8.32$, $\epsilon=1.60$, RMSE $=0.05$,
    (f) $P[A]=(1+ps^{n_A-n_B})^{-1}$, $p=0.81$, $s=0.75$, RMSE $=0.04$.
    The different colors correspond to $n_B=0$ (black), 1 (red), 2 (green), 3 (blue), 4 (yellow), 5 (brown), 6 (grey), 7 (violet).}
\label{f:fittings1}
 \end{figure}

\begin{figure}[t]
\centerline{\includegraphics[width=0.5\textwidth]{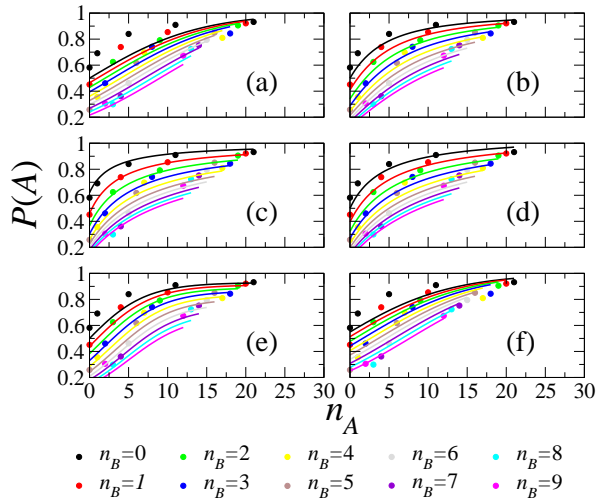}}
\caption{
{\bf Testing different models for data set $D_2$.}
(a) $P[A]=\delta^{n_A}/(\delta^{n_A}+\delta^{n_B})$, $\delta=1.15$, RMSE $=0.07$,
(b) $P[A]=(\delta+n_A)^{\epsilon}/[(\delta+n_A)^{\epsilon}+(\delta+n_B)^{\epsilon}]$, $\delta=3.66$, $\epsilon=1.52$, RMSE $=0.05$,
(c) $P[A]=(\delta+\epsilon n_A)/[1+\epsilon(n_A+n_B)]$, $\delta=0.62$, $\epsilon=0.35$, RMSE $=0.05$,
(d) $P[A]=1/2+\delta(n_A-n_B)/(n_A+n_B+\epsilon)$, $\delta=0.57$, $\epsilon=4.39$, RMSE $=0.06$,
(e) $P[A]=[1+(1+\delta\epsilon^{-n_A})/(1+\delta\epsilon^{-n_B})]^{-1}$, $\delta=12.01$, $\epsilon=1.30$, RMSE $=0.05$,
(f) $P[A]=(1+ps^{n_A-n_B})^{-1}$, $p=0.82$, $s=0.87$, RMSE $=0.06$.
The different colors correspond to $n_B=0$ (black), 1 (red), 2 (green), 3 (blue), 4 (yellow), 5 (brown), 6 (grey), 7 (violet), 8 (cyan), 9 (pink).}
\label{f:fittings2}
\end{figure}

In the literature one can find different models that propose different functional forms for $P[A|n_A,n_B]$. Following \cite{Arganda2012PNAS}, we fitted several of them \cite{Arganda2012PNAS,Perez-Escudero2011PlosComputBiol,Goss1989Naturw,Deneubourg1990JInsectBehav,Meunier2006Behav} to the current data. The quality of fitting is shown in Figs.~\ref{f:fittings1} and \ref{f:fittings2} for data sets $D_1$ and $D_2$, respectively, in different colors for different values of $n_B$. For $D_1$, the best results are produced with the model in Ref.~\cite{Meunier2006Behav} with a RMSE $\approx 0.035$, followed closely by the model presented in
this paper (RMSE $\approx 0.042$). For $D_2$, the best fitting (RMSE $\approx 0.046$) is produced with the model in Ref.~\cite{Meunier2006Behav}, and followed closely by the models in Refs.~\cite{Goss1989Naturw,Deneubourg1990JInsectBehav} (RMSE $\approx 0.053$) and Ref.~\cite{Arganda2012PNAS} (RMSE $\approx 0.054$), with none of them being the one in this paper. See Table~\ref{t:fittings} for more information.

It should be noted that the first model in Table~\ref{t:fittings} is equivalent to a special case of our model (i.e., $p=1$). Therefore, the fitting cannot be better than the present model. Note also that we fitted the model in Ref.~\cite{Arganda2012PNAS} with $k=0$. The result of fitting with $k$ as a free parameter gives rise to very small values of $k$ ($k=0.04$ for $D_1$ and $k=0.065$ for $D_2$), in the order of $10^{-2}$. The parameter $\epsilon$ is insensitive to the small value of $k$ being different to $0$ ($\epsilon=1.60$ for $D_1$ and $\epsilon=1.30$ for $D_2$), while parameter $\delta$ is a much more sensitive ($\delta=7.04$ for $D_1$ and $\delta=8.72$ for $D_2$) (compare to results in Table~\ref{t:fittings}), as the minimum in the optimization is flatter in the direction of the $\delta$ parameter, as happens also for parameter $p$ in the present model (see Fig.\ref{f:fittingerr}). The quality of the fittings is of the same order as when using $k=0$ (RMSE $=0.054$ for $D_1$ and RMSE $=0.054$ for
$D_2$). This
also happens for the zebrafish data in Ref.~\cite{Arganda2012PNAS}.

% We only support three levels of headings, please do not create a heading level below \subsubsection.
% \subsection*{Subsection 1}
%
% \subsubsection*{SubSubsection 1.1}
%
% \subsection*{Subsection 2}

%%%%%%%%%%%%%%%%%%%%%%%%%%%%%%%%Fig 5

%%%%%%%%%%%%%%%%%%%%%%%%%%%%%%%%
\section{Discussion}

We showed that the simple Bayesian model provides a quantitative agreement with behavioral data of humans sequentially answering questions with binary options.
At least two other studies used the same model as ours to be fit to data in different
contexts. In Ref.~\cite{Perez-Escudero2011PlosComputBiol},
sequential choices by fish between two identical refugia are
modeled. Depending on whether the two refugia are identical or
nonidentical (i.e., only one arm was with a replica predator), the
unbiased prior ($p=1$ in our notation) or a biased one ($p\neq 1$) is used,
respectively. In both unbiased and biased prior cases, the authors
concluded $s\approx 0.4$ (and the results are robust for $0.25\le s\le
0.5$), translating into $c=1/(s+1)\approx 0.7$ in our notation.
In another experiment
with a different fish species, where fish individuals chose one of the
two arms of a maze to avoid replica predators, Ward and colleagues
\cite{Ward2011PNAS} estimated $s\approx 1/e^{0.478}\approx 0.62$,
translating into $c\approx 0.62$.  In contrast, our results indicate
$s\approx 0.7-0.8$ and hence $c\approx 0.56-0.59$. This difference
may result from different species; humans may have lower
responsitivity to social stimuli (i.e., $c$ value closer to 0.5) than
fish (see Ref.~\cite{Traulsen2010PNAS_exper} for related experiments). The type of the task may also contribute to this difference. In the current study, the data set used are quizzes asking general knowledge of the participants. By contrast, in the fish experiments, the binary choice between two pathways that were identical except for the possible presence of a replicator predator was made by fish.

%%%%%%%%%%%%%%%%%%%%%%%%%%%%%%%%Fig 6

Quantitatively, some models fit better to our data than the present model does, in particular for data set $D_2$ (Table~\ref{t:fittings}). However, it should be noted that some of these previous models were proposed as fits, without particular mechanistic derivation \cite{Goss1989Naturw,Deneubourg1990JInsectBehav,Meunier2006Behav}. Another model, i.e., the fourth model in Table~\ref{t:fittings} \cite{Arganda2012PNAS}, which results from the Taylor expansion of the model proposed in Ref.~\cite{Perna2012PlosComputBiol}, has mechanistic underpinning. However, the model is derived from ant's random walk on a specific arena \cite{Perna2012PlosComputBiol}. In particular, the exit point that corresponds to the decision of one of the two alternatives is literally the spatial exit point of the animal. That may be why this model \cite{Arganda2012PNAS,Perna2012PlosComputBiol} does not fit well to the present data. Compared to Arganda's model \cite{Arganda2012PNAS} (fifth model in Table~\ref{t:fittings}), the present
model fits better to data set $D_1$ and worse to $D_2$.

A way to differentiate between models is to have data on the behavior for large number of information sources (large $r$). In that limit the different models provide different functional forms for $P[x=A]$. Therefore, the models from Table~\ref{t:fittings} give rise to different limits $r\rightarrow\infty$. The first and the last one (model used in this paper) give rise to a step function. The second model converges to $x^{\epsilon}/(x^{\epsilon}+(1-x)^{\epsilon})$, where $x$ is the fraction of A responses, which coincides with Weber's law for $\epsilon=1$. However, the values of $\epsilon$ estimated for our data are much larger than unity. The third function for large $r$ approximates the fraction of A responses. The fourth function gives $1/2+\delta (2x-1)$, which is a good approximation of the previous model given that the fitting parameter $\delta\simeq1/2$ for our datasets. The fifth model gives a constant value $P[x=A]=1/2$ in the limit $r\to\infty$. More experimental data for large $r$ would enable
the further validation of
models.

\begin{table*}[!ht]
  \caption{\bf{Fitting results for different models}}
  \centering
  \begin{tabular}{c|c|c|c|c|c|}
  Model $P[A]$ & \multicolumn{4}{c|}{Fitted parameters and RMSE}& [Refs.] \\
  \cline{2-5}
  & $D_1$ & RMSE & $D_2$ & RMSE & \\ \hline \hline
  $\frac{\delta^{n_A}}{\left(\delta^{n_A}+\delta^{n_B}\right)}$ & $\delta=1.33$ & $0.061$ & $\delta=1.15$ & $0.070$ & \cite{Perez-Escudero2011PlosComputBiol} \\ [10pt]\hline
  $\frac{(\delta+n_A)^{\epsilon}}{(\delta+n_A)^{\epsilon}+(\delta+n_B)^{\epsilon}}$ & $\delta=3.90$, $\epsilon=1.95$ & $0.054$ & $\delta=3.66$, $\epsilon=1.52$ & $0.053$ & \cite{Goss1989Naturw,Deneubourg1990JInsectBehav} \\ [10pt]\hline
  $\frac{\delta+\epsilon n_A}{1+\epsilon(n_A+n_B)}$ & $\delta=0.63$, $\epsilon=0.41$ & $0.035$ & $\delta=0.62$, $\epsilon=0.35$ & $0.046$ & \cite{Meunier2006Behav} \\ [10pt]\hline
  $\frac{1}{2}+\delta\frac{n_A-n_B}{n_A+n_B+\epsilon}$ & $\delta=0.68$, $\epsilon=5.06$ & $0.056$ & $\delta=0.57$, $\epsilon=4.39$ & $0.057$ & \cite{Perna2012PlosComputBiol} \\ [10pt]\hline
  $\left(1+\frac{1+\delta\epsilon^{-n_A}}{1+\delta\epsilon^{-n_B}}\right)^{-1}$ & $\delta=8.32$, $\epsilon=1.60$ & $0.054$ & $\delta=12.01$, $\epsilon=1.30$ & $0.054$ & \cite{Arganda2012PNAS} \\ [10pt]\hline
  $(1+ps^{n_A-n_B})^{-1}$ & $p=0.81$, $s=0.75$ & $0.042$ & $p=0.82$, $s=0.87$ & $0.059$ & [here] \\ [10pt]\hline
  \end{tabular}
  \begin{flushleft}Results of fitting different models for $P[A]$ to data sets $D_1$ and $D_2$.
  \end{flushleft}
  \label{t:fittings}
\end{table*}

There are some limitations of the present study. First, we ignored the individuality of the respondents. In fact, for each question, there should be those who know the correct answer and those who do not.
Such personal knowledge can be incorporated to models for sequential answering
 \cite{Banerjee1992QuarterlyJEcon,Mori2012PhysRevE}. Clarifying this issue warrants future work. Second, we tried to incorporate the information about the previous responses into our model. However, the design of the experiment makes it difficult to cope with this issue. The answers offered to subject $i$ in each round are not a random sample from the pool of responses in the previous round, but are the responses of the previous respondents $i-1, i-2, ..., i-r$ as initially labeled, which represents a biased sampling. Together with the influence of the history of self-responses on the new decision, these features affect the decision making process of the subjects and thus the evolution of the fraction of correct answers. Indeed in many situations individuals are not making decision from a tabula rasa but they are shaping decisions continuously from social interactions and external signals. Future developments of the theory are expected to incorporate these ingredients to deal with more realistic situations.
Besides, large scale experiments taking advantage of the new technologies available would be welcome to confront with decision making theories.

% Do NOT remove this, even if you are not including acknowledgments.

\section*{Acknowledgments}

We acknowledge the authors of Refs.~\cite{Mori2012PhysRevE,Mori2013JPhysSocJapan} for making their data open to public. We also thank Shintaro Mori for discussion and giving us information about the detailed procedure of their data acquisition. We also acknowledge insightful discussions with Gonzalo G. de Polavieja and Konstantin Klemm.

% \section*{References}

% \bibliography{baydmJFG}

\end{document}